\documentclass[12pt]{article}

\font\twelvemsb=msbm10 scaled 1200 
\def\Bbb#1{\hbox {\twelvemsb#1}}

\newcommand\bm[1]{\mbox{\boldmath$#1$}}
\newcommand\cm[1]{\bm{\cal#1}}

\newcommand{\C}{{ \cal C}} 
 
\newcommand\ov[1]{\overline{#1}} 
\newcommand{\F}{{\cal F}}  
\newcommand{\K}{{\cal K}}

\newcommand{\I}{{\cal I}}
\newcommand{\Fsq}{\bm{{\cal F}^2}}

\begin{document}

\title{The Wahlquist-Newman solution}
\author{Marc Mars\thanks{Also at Laboratori de F\'{\i}sica Matem\`atica,
Societat Catalana de F\'{\i}sica, IEC, Barcelona,Spain.}\\ 
Albert Einstein Institut \\
Am M\"uhlenberg 1, D-14476 Golm, Germany.}
\date{5 January 2001}
\maketitle
\begin{abstract}
Based on a geometrical property which holds both for the Kerr metric and
for the Wahlquist metric we argue that the Kerr metric
is a vacuum subcase of the Wahlquist perfect-fluid solution.
The Kerr-Newman metric is a physically preferred
charged generalization of the Kerr metric. We discuss which 
geometric property makes this metric so special and claim that
a charged generalization of the Wahlquist metric satisfying a
similar property should exist. This is the Wahlquist-Newman metric,
which we present explicitly in this paper. This family of metrics has eight
essential parameters and contains the Kerr-Newman-de Sitter and the Wahlquist
metrics, as well as the whole Pleba\'nski limit of the
rotating C-metric, as particular cases.
We describe the basic geometric properties of the Wahlquist-Newman
metric, including the electromagnetic field and its sources,
the static limit of the family and the extension of the spacetime across
the horizon.
\end{abstract}


\newpage

\section{Introduction}

Among the few explicitly known stationary (non-static) and axially
symmetric perfect-fluid spacetimes, the Wahlquist family
\cite{Wahlquist} enjoys a privileged position. First, it is the
oldest known solution and it remains, in some sense,
the simplest one. More importantly, it has interesting physical properties
(see \cite{2Wahlquist} and references therein) which have made
this metric a good candidate to describe the interior
of an isolated rotating body in equilibrium. This view has
been recently challenged in \cite{ImpoWahl}, where 
the matching conditions between the Wahlquist
metric and a vacuum, asymptotically flat spacetime are claimed
to be incompatible in a perturbative sense. This strongly
indicates that the Wahlquist metric does not describe
the interior or a rotating body in
vacuum. In order to make this result conclusive it would be of interest
to develop a proper theoretical
analysis of the perturbative approach to the matching conditions.

In any case, the fundamental properties which make
the Wahlquist metric so special are of geometrical nature.
Indeed, this metric is known to be uniquely characterized
among stationary, rigidly rotating,  perfect-fluid spacetimes 
by any of the following, seemingly unrelated, properties
(see \cite{SenoinChin} for a discussion),
\begin{itemize}
\item[(1)] The Simon tensor vanishes \cite{Kramer}.
\item[(2)] The spacetime admits a Killing tensor of type [(11)(11)].
\cite{Papa}
\item[(3)] The spacetime is axially symmetric, the Weyl tensor is 
Petrov type D and the equation of state of the perfect fluid is
$\rho + 3 p = \mbox{const.}$ \cite{Seno}.
\end{itemize}
For the purposes of this paper, characterization (1) will be
the most relevant one. The Simon tensor \cite{Simon} was 
put forward in order
to obtain a unique characterization of  the Kerr metric \cite{Kerr}.
More precisely, 
the Kerr spacetime is the only strictly stationary (i.e. with
a Killing vector which is timelike everywhere), vacuum and
asymptotically flat spacetime for which the Simon tensor vanishes. 
This fact, combined with (1), shows that
there may exist a close relationship between the Wahlquist metric and
the Kerr metric. However, no such relationship has been found so far.
One of the aims of this paper is to show that the Kerr metric can
be obtained as a particular, vacuum, subcase of the Wahlquist metric.
In fact, we will also show that the Kerr-de Sitter
metric \cite{Carter}, which is vacuum with a cosmological
constant, belongs to the Wahlquist family in the limit
$\rho + p =0$.

The existence of physically privileged charged generalization of the Kerr and
Kerr-de Sitter metrics, namely the Kerr-Newman \cite{KNewman}
and Kerr-Newman-de Sitter spacetimes\cite{Carter}, leads us to consider
whether a similar, privileged,
charged generalization of the Wahlquist metric exists. To analyze such 
a question we should first make precise the meaning of the term
``privileged''. As we shall see, the Kerr, the Kerr-de Sitter
and the Kerr-Newman-de Sitter metrics have very special geometric
properties which relate the Weyl tensor, the  Killing vector and the
electromagnetic field (when one is present). Moreover, these
conditions turn out to be fulfilled also by the Wahlquist metric.
Thus, there exists a geometrically clear sense
in which a privileged charged generalization of
the Wahlquist metric might exist. We call it Wahlquist-Newman
metric, first because it contains both Wahlquist and Kerr-Newman-de Sitter 
as particular cases and also in order to emphasize the very special
geometrical properties fulfilled by this spacetime.
The main objective of this paper is to obtain the explicit form
of this metric.  It turns out that the Wahlquist-Newman
family contains eight arbitrary parameters.  It
represents a rigidly rotating perfect fluid, which may be
charged or not, together with an electromagnetic field. The
sources of the electromagnetic field are the perfect fluid (when this
is charged) and/or a singularity of the spacetime. 
The latter corresponds to the singular source hidden behind the
event horizon in the Kerr-Newman spacetime. 

The paper is organized as follows. In section 2, we recall the
relationship between the vanishing
of the Simon tensor and  the Weyl tensor and we discuss
which geometrical properties make the Kerr, Kerr-Newman,
Kerr-Newman-de Sitter and Wahlquist metrics so special. 
In section 3, we rewrite the Wahlquist metric in such a way
that the Kerr-de Sitter metric (and the Kerr metric) are contained 
as particular subcases. 
In section 4, we present the Wahlquist-Newman metric
and we describe its fundamental properties. 
First, we stress that the 
geometrical properties described in section 2  also hold
for this metric. Then, we give
the explicit expressions for the energy-density, the pressure and 
the fluid velocity of the perfect
fluid. The  electromagnetic field and its charge current are also 
written down
and the number of essential parameters in the family is discussed.
We also show that the particular case in which the perfect fluid
vanishes  corresponds to the well-known Pleba\'nski metric \cite{Ple},
which is an important limiting case of the rotating C metric \cite{PlDe}. This
shows, in particular, that the Wahlquist-Newman spacetime contains
the Kerr-Newman-de Sitter metric as a particular case.
In section 5, we analyze the 
static limit of the Wahlquist-Newman metric. To do that, we rewrite
the metric in a suitable coordinate system which admits an explicit static
limit and which, in addition, allows
for an extension of the Wahlquist-Newman spacetime
across its horizon
(although the metric represents a perfect fluid, it does have a regular
horizon, as we shall see). Finally, we include an Appendix where
Einstein-Maxwell's equations under the assumptions of this paper are solved.

\section{Geometric properties of the Wahlquist and the
Kerr-Newman-de Sitter metrics}
\label{Prop}

The Kerr metric and the Wahlquist metrics share the property
that the Simon tensor \cite{Simon} vanishes identically.
The geometrical meaning of the vanishing of the Simon tensor 
in vacuum has been recently clarified in \cite{Mars1}. The fundamental
underlying property is a close relationship between
the Weyl tensor and the stationary Killing
vector. Properties of the Weyl tensor can be quite naturally described
using the language of self-dual
two forms, which are two-forms $\cm{X}$ satisfying
$\cm{X}^{\star} = - i \cm{X}$ where $\star$ denotes the Hodge dual with
respect to the volume form $\eta_{\alpha\beta\gamma\delta}$. 
From the Weyl tensor $C_{\alpha\beta\gamma\delta}$
and the stationary Killing vector $\vec{\xi}$ we can write
down two canonical self-dual objects, the self-dual
Weyl tensor $\C_{\nu\mu\alpha\beta} \equiv
C_{\nu\mu\alpha\beta} +
\frac{i}{2}  \eta_{\alpha\beta\rho\sigma} C_{\nu\mu}^{\,\,\,\,\,\,\,\,
\rho\sigma}$ and the so-called Killing form 
$\F_{\alpha\beta} \equiv \nabla_{\alpha} \xi_{\beta} + \frac{i}{2}
\eta_{\alpha\beta\gamma\delta} \nabla^{\gamma} \xi^{\delta}$. 
It is natural to ask which spacetimes have the property that
the self-dual Weyl tensor and the Killing form are related to each other.
The simplest relationship between these two objects
which respects all the symmetries of the self-dual Weyl tensor is
(the object 
$\I_{\alpha\beta\gamma\delta} \equiv
(g_{\alpha\gamma}\, g_{\beta\delta}-g_{\alpha\delta}\,g_{\beta\gamma}
 + i \,\eta_{\alpha\beta\gamma\delta})/4$ is the canonical metric
in the space of two-forms)
\begin{eqnarray}
\C_{\alpha\beta\gamma\delta} =
L \left (\F_{\alpha\beta} \F_{\gamma\delta} - \frac{1}{3}
\I_{\alpha\beta\gamma\delta} \Fsq \right ), 
\label{weylSi}
\end{eqnarray}
where $L$ is a complex, scalar function 
and $\Fsq \equiv \F_{\alpha\beta} \F^{\alpha\beta}$. It turns out \cite{Mars1}
that the vanishing
of the Simon tensor in vacuum is equivalent to  (\ref{weylSi}).
We know that the Simon tensor vanishes for the Wahlquist metric. So, 
we can ask whether (\ref{weylSi}) holds also 
for the Wahlquist spacetime. A straightforward calculation shows that this is
indeed the case. Actually, it can be seen that 
the original assumptions
made by Wahlquist in order to find his spacetime, although written
in another formalism (see \cite{2Wahlquist}), can be rewritten so that
they consist of condition (\ref{weylSi}) plus axial
symmetry. Thus, with hindsight, Kramer's uniqueness result \cite{Kramer}
of the Wahlquist metric is equivalent to dropping the condition of
axial symmetry from Wahlquist's original assumptions.

Following the discussion in the Introduction, we can ask whether
condition (\ref{weylSi}) is also fulfilled by Kerr-de Sitter,
Kerr-Newman and Kerr-Newman-de Sitter. The answer is yes, as
a simple calculation shows. However, the Kerr-Newman and the
Kerr-Newman-de Sitter spacetimes contain, in addition, an electromagnetic
field. So we should analyze whether this field
fits nicely into the geometrical relation (\ref{weylSi}). This
is very important for our purposes because it will determine
what makes these charged spacetimes so special, and it will
indicate how the  charged generalization of Wahlquist metric should be defined.
Let us call the electromagnetic field as $K_{\alpha\beta}$. 
This two-form defines canonically a self-dual two-form according to
$\K_{\alpha\beta} \equiv K_{\alpha\beta} + i K^{\star}_{\alpha\beta}$.
It can be easily checked that in Kerr-Newman and Kerr-Newman-de Sitter
the self-dual electromagnetic field  {\it
is proportional} to the Killing form, i.e. $\K_{\alpha\beta}
\propto \F_{\alpha\beta}$. This is the most natural relationship 
one could think of between these two objects. Thus, all these
metrics do have very special geometrical properties.

This discussion above indicates two things.
First, that the Wahlquist metric is likely to contain
the Kerr-de Sitter metric (and hence the  Kerr metric)
as a particular subcase and, second,
that a charged generalization of
Wahlquist should also exist satisfying the following properties:
1) It contains both Wahlquist 
and Kerr-Newman-de Sitter as subcases, 2) it satisfies the
relationship
(\ref{weylSi}) between the Weyl tensor and the Killing form and 3)
its self-dual electromagnetic field  is proportional to the Killing form.
Its energy-momentum tensor
should contain both an electromagnetic field part and a perfect-fluid part.

Table 1 shows graphically the interrelationships between these metrics.
Single arrows indicate well-established and natural generalizations
and arrows between question marks indicate plausible
relations between metrics.
In particular, it becomes apparent  that some metric,
the Wahlquist-Newman metric, should fill  the lower, right corner of
this table.

\vspace{5mm}

\begin{center}

\begin{tabular}{|ccc|}
\hline
 {\bf Non-charged metrics} & & 
$\frac{}{}${\bf Charged counterparts} \\
\hline 
& & \\
$\frac{}{}$ Kerr & $\longrightarrow$ & $\frac{}{}$ Kerr-Newman  \\
{\Large $\downarrow$} & & {\Large $\downarrow$} \\
$\frac{}{}$ Kerr-de Sitter & $\longrightarrow$ & 
$\frac{}{}$ Kerr-Newman-de Sitter \\
\hspace{2mm} ? {\Large $\downarrow$ } ?  & & 
\hspace{2mm} ? {\Large  $\downarrow$ } ?  \\
$\frac{}{}$ Wahlquist & $? \longrightarrow$ ?&  
\framebox{$\frac{}{}$\mbox{Wahlquist-Newman?}} \\
& & \\
\hline
\end{tabular}

\noindent Table 1: Relationships between the metrics discussed in this paper
\end{center}

\section{Kerr-de Sitter limit in the Wahlquist family}

Let us start  by writing down the line-element of the Wahlquist family
as it appears in \cite{2Wahlquist}. This is actually a generalization 
(by adding a discrete parameter) of the
original Wahlquist metric as given in 
\cite{Wahlquist} and was originally given by Senovilla
in \cite{Seno}  (see \cite{MS} for a discussion on
the different published versions of the Wahlquist metric 
 and their interrelationships). The
Wahlquist line-element is
\begin{eqnarray}
ds^2 = -\frac{1}{\Phi^2} \left ( dt - A d\theta \right )^2 + r^2 d\theta^2
+ \frac{g}{\mu_0} \left ( \frac{du^2}{h_1} + \frac{dv^2}{h_2} \right ),
\end{eqnarray}
where 
\begin{eqnarray*}
h_1(u) & = & h_0 + \epsilon_0 \cos \left (2 u \right ) + \left (u+u_0
\right ) \sin \left (2 u \right ), \\
h_2(v) & = & h_0 -\epsilon_0 \cosh \left (2 v \right ) + \left (v+v_0
\right ) \sinh \left (2 v \right ), \\
g & = & \cos \left (2u \right ) + \cosh \left (2v \right ), \hspace{1cm}
\frac{1}{\Phi^2} = \frac{h_1 - h_2}{\kappa g}, \hspace{1cm}
r^2 = 4 r_0^2 \Phi^2 h_1 h_2, \\
A & = & - 2 \kappa r_0 \cosh \left(v_A \right ) + \frac{2 \kappa r_0
\left ( h_2 \cos \left (2u \right ) + h_1 \cosh \left (2v \right )
\right )}{h_1-h_2}.
\label{Wahl}
\end{eqnarray*}
All symbols with zero subscripts, as well as $\kappa$ and $v_A$, are
arbitrary constants.
The energy-momentum of this spacetime is 
a rigidly rotating perfect fluid (i.e. its velocity  vector is
proportional to the Killing  vector $\vec{\xi}= \partial_t$).
The energy-density $\rho$ and pressure $p$
are $\rho = \mu_0 \left ( 1 - \frac{\kappa}{\Phi^2}\right )$ and
$p = \mu_0 \left ( 3 \frac{\kappa}{\Phi^2} -1 \right )$, so that
the equation of state is $\rho + 3 p = 2 \mu_0$. We want to
rewrite this metric in such a way that the Kerr metric is included as
a particular case.
We first rescale $u$ and $v$ as follows, $u = \beta y + \pi/2$, $v = \beta z$,
where $\beta$ is any non-zero constant.
The  function $g$ transforms
into $g = \cosh \left (2 \beta z \right ) - \cos \left (2 \beta y \right )$.
The constant $\beta$ is superfluous as long as it
remains non-zero, but it may be that the limit $\beta \rightarrow
0$ gives another metric, perhaps the Kerr metric we are
looking for. In order to work out this idea, we should  
make $\beta \rightarrow 0$ meaningful. This requires some 
redefinitions of constants. We start by defining
\begin{eqnarray}
Q(y,z) \equiv \frac{\cosh \left (2 \beta z \right ) - \cos \left (
2\beta y \right )}{2 \beta^2},
\label{defQ}
\end{eqnarray}
which is regular at $\beta = 0$. The $2 \times 2$ block spanned by
$\left \{u,v\right \}$ in (\ref{Wahl}) takes the form
$Q \left ( U(z)^{-1} dz^2 + V(y)^{-1} dy^2 \right )$,
where $U = \mu_0 h_2 /(2 \beta^4)$ and $V = \mu_0 h_1/(2 \beta^4)$.
The constants must be redefined so that $U$ and $V$ are regular
at $\beta =0$. Furthermore $\mu_0$ should remain finite
and non-zero (because of the relation
$\rho + 3 p = 2 \mu_0$, which is non-zero in the Kerr-de Sitter metric).
In addition, the number 
of parameters should not be reduced in the limit $\beta \rightarrow 0$.
All this is achieved by the following redefinition of constants
\begin{eqnarray}
\mu_0 \mbox{ invariant}, \hspace{1cm}
\frac{h_0 \mu_0}{2 \beta^4} = Q_0 + \frac{\nu_0}{2 \beta^2} +
\frac{\mu_0}{2 \beta^4},
\hspace{1cm}
\frac{\mu_0 \epsilon_0}{\beta^2} = \nu_0 + \frac{\mu_0}{\beta^2}, 
\nonumber \\
\frac{\mu_0 v_0}{\beta^3} = a_1, \hspace{1cm}
\frac{\mu_0 \left (u_0 + \pi/2 \right)}{\beta^3} = - a_2,
\hspace{3cm}
\label{redef}
\end{eqnarray}
which brings $U$ and $V$ into the form
\begin{eqnarray}
U = Q_0 - \frac{\mu_0}{2 \beta^2}
\left [ \frac{\cosh \left (2 \beta z \right ) -1 }{\beta^2} -
\frac{z \sinh \left ( 2\beta z \right )}{\beta} \right ]
+ \nu_0 \frac{1 - \cosh \left ( 2 \beta z \right )}{2 \beta^2} +
a_1 \frac{\sinh \left (
2 \beta z \right )}{2 \beta}, \nonumber \\
V = Q_0 + \frac{\mu_0}{2 \beta^2}
\left [ \frac{1- \cos \left (2 \beta y \right ) }{\beta^2} -
\frac{y \sin \left ( 2\beta y \right )}{\beta} \right ]
+ \nu_0 \frac{1 - \cos \left ( 2 \beta y \right )}{2 \beta^2} +
a_2 \frac{\sin \left (
2 \beta y \right )}{2 \beta}.
\label{UV}
\end{eqnarray}
We should now analyze the $\{t,\theta \}$ block.
The constants $\kappa$,
$r_0$ and $v_A$ correspond to the freedom of
performing linear coordinate changes in 
$t$ and $\theta$. Since the coordinates should remain adapted to
the Killing vector $\vec{\xi}$ (which is privileged both
for the Wahlquist and for the Kerr metrics), we consider changes of the
type  $\tau = b_1 \left ( t + b_2
\theta \right ), \sigma = b_3 \theta$. Let us choose
\begin{eqnarray*}
b_1 = \frac{\beta}{\sqrt{\mu_0 \kappa}}, \hspace{1cm}
b_2 = 2 \kappa r_0 \left ( \cosh \left (v_A \right ) -1 \right )
, \hspace{1cm}
b_3 = 4 \beta^3 r_0\sqrt{ \frac{\kappa}{\mu_0} },
\end{eqnarray*}
which bring the  Wahlquist line-element (\ref{Wahl}) into the
form
\begin{eqnarray}
ds^2 = - \lambda \left ( d \tau - \frac{ v_1 V + v_2 U}{V - U} 
d\sigma \right )^2 + \frac{U V}{\lambda} d\sigma^2 +
\left (v_1 + v_2 \right )
\left ( \frac{dy^2}{V} +
\frac{dz^2}{U} \right ),
\label{Wahl2}
\end{eqnarray}
where $U$ and $V$ are given by (\ref{UV}), $v_1$, $v_2$ read
\begin{eqnarray}
v_1 = \frac{ \cosh \left (2 \beta z \right ) - 1 }{2 \beta^2}, 
\hspace{1cm} 
v_2 = \frac{1 - \cos \left (2 \beta y \right )}{2 \beta^2},
\label{v1v2}
\end{eqnarray}
and $\lambda = (V-U)/(v_1 + v_2)$. All metric functions in 
(\ref{Wahl2}) are independently regular at $\beta =0$. 
The structure of this line-element is very similar to the one given by
Senovilla in  \cite{Seno}, the only difference being the choice of
parameters. It is not difficult to obtain the redefinitions which bring
Senovilla's form into (\ref{Wahl2}). Thus, a regular limit $\beta =0$
could also have been obtained starting from that line-element. We
preferred to start from (\ref{Wahl}) in order to deal only with
essential parameters.

The explicit form of the metric (\ref{Wahl2})
when  $\beta =0$ is (after trivially reorganizing
the block $\{\tau,\sigma \}$) 
\begin{eqnarray}
ds^2 = - \frac{\hat{V}}{y^2+z^2} \left (d\tau - z^2 d\sigma \right )^2
+ \frac{\hat{U}}{y^2+z^2} \left (d\tau + y^2 d\sigma \right )^2 +
\left (y^2 + z^2\right ) \left ( \frac{dy^2}{\hat{V}} +
\frac{dz^2}{\hat{U}} \right ).
\end{eqnarray}
where 
\begin{eqnarray*}
\hat{U} = Q_0 + \frac{\mu_0}{3} z^4 - \nu_0 z^2 + a_1 z, \hspace{1cm}
\hat{V} = Q_0 + \frac{\mu_0}{3} y^4 + \nu_0 y^2 + a_2 y, \\
\hat{\lambda} = \nu_0 + a_2 \frac{y}{y^2 + z^2} - a_1 \frac{z}{y^2 + z^2}
+ \frac{\mu_0}{3} \left ( y^2 - z^2 \right ).
\end{eqnarray*}
This is the uncharged subcase of the Pleba\'nski metric \cite{Ple},
which is an important limiting case of the Pleba\'nski-Demia\'nski metric,
also called rotating C-metric \cite{PlDe}.
The constant $a_1$ is
closely related to the NUT-parameter and $a_2$ is related
to the mass parameter. A particular case of this metric is obtained by
setting $Q_0 = a^2$, $\nu_0=1 - a^2 \Lambda/3$, $a_1=0$ and
redefining $a_2 \rightarrow -2 M$
and $\mu_0 \rightarrow - \Lambda$. After
the coordinate changes 
\begin{eqnarray}
y = r, \hspace{1cm} z = a \cos \theta, \hspace{1cm} a \sigma =
\frac{- \phi}{1 + \frac{1}{3} \Lambda a^2},
\hspace{1cm} \tau = \frac{t - a \, \phi}{1 + \frac{1}{3} \Lambda a^2},
\label{coordchange}
\end{eqnarray}
we obtain the Kerr-de Sitter metric \cite{Carter} in 
Boyer-Lindquist coordinates. 
\begin{eqnarray}
ds^2 = \rho^{-2} \left [ - \Delta_r \bm{\alpha_0}^2
+  \Delta_{\theta} \sin^2 \theta \,\,
 \bm{\alpha_1}^2 \right ]
+ \rho^2 \left ( \frac{dr^2}{\Delta_r} + \frac{d\theta^2}{\Delta_{\theta}}
\right ),
\label{KerrdeSitter}
\end{eqnarray}
where $\rho^2 = r^2 + a^2 \cos^2 \theta$, $\Delta_r = \left (a^2
+ r^2 \right )\left ( 1 - \frac{1}{3} \Lambda r^2 \right ) - 2 M r$
and $\Delta_{\theta} = 1 + \frac{1}{3} \Lambda a^2 \cos^2 \theta$.
The one-forms $\bm{\alpha_0}$ and $\bm{\alpha_1}$ are
\begin{eqnarray*}
\bm{\alpha_0} =\frac{1}{1 + \frac{1}{3} \Lambda a^2} \left ( dt - a 
\sin^2 \theta d\phi \right ), \hspace{1cm}
\bm{\alpha_1} = \frac{1}{1 + \frac{1}{3} \Lambda a^2} \left [
a \, dt - \left ( a^2 + r^2 \right ) d\phi  \right ].
\end{eqnarray*}
Of course, by setting $a=0$ in this metric
we get the Schwarzschild-de Sitter metric and the particular case
$\Lambda=0$ is the Kerr metric. Thus, 
the Wahlquist metric {\it does} contain Kerr-de Sitter (and Kerr) 
as a particular case, as we wanted to prove.

\section{The Wahlquist-Newman family of metrics}
\label{WN}

The line-element of the Kerr-Newman-de Sitter spacetime
can be obtained from (\ref{KerrdeSitter}) just by modifying 
the function $\Delta_r$ with an additive constant, i.e.
\begin{eqnarray*}
\Delta_r = \left (a^2 + r^2 \right ) \left (1- \frac{1}{3} \Delta r^2
\right ) - 2 M r + q^2,
\end{eqnarray*}
the constant $q$ being directly related to the electric charge of 
the black hole (and hence to the electromagnetic field).
By analogy, we assume as a working hypothesis
that the Wahlquist-Newman metric we are seeking can be obtained
by modifying the
functions on the block $\{dy,dz\}$ of the metric (\ref{Wahl2}). 
The reason why we must allow 
both functions $V(y)$ and $U(z)$ to be changed instead of only one 
(as in the Kerr-Newman-de Sitter case) will become clear later.
So, let us assume that the
Wahlquist-Newman metric can be written in the form
\begin{eqnarray}
ds^2 = - \frac{V_1}{v_1 + v_2} \left ( d \tau - v_1 d\sigma \right )^2
+ \frac{U_1}{v_1 + v_2} \left ( d\tau + v_2 d\sigma \right )^2 + \left (v_1 + 
v_2 \right ) \left ( \frac{dy^2}{V_1} + \frac{dz^2}{U_1} \right ),
\label{Wahlgen}
\end{eqnarray}
where $V_1(y)$ and $U_1(z)$ are unknown functions and $v_1(z)$, $v_2(y)$
are given by (\ref{v1v2}).
We want to solve the Einstein-Maxwell field equations for an energy-momentum
tensor $T_{\mu\nu}$ consisting of two parts: a perfect-fluid component
$T_{\mu\nu}^{pf}$
with the fluid velocity being proportional to the stationary Killing vector
$\vec{\xi} = \partial_{\tau}$ and an
electromagnetic part $T_{\mu\nu}^{em}$. If we denote
by $K_{\mu\nu}$ the electromagnetic field 
and by ${\cal K}_{\mu\nu}$ its self dual part, we want to 
impose  $\cm{K} \propto \cm{F}$, so that the fundamental geometric property
satisfied by the Kerr-Newman metric is preserved.  In Kerr-Newman,
the electromagnetic field is source-free (more precisely, the source
of the electromagnetic field is located at the singularity inside the black
hole). This is most reasonable because there is no matter to
support  electric charge. In our case, however, there is a perfect fluid 
which may perfectly be charged. So, we admit 
a charge current $\vec{j}$ proportional to the fluid velocity $\vec{u}$.
Hence, Maxwell's equations read
\begin{eqnarray}
d \bm{K} = 0,\hspace{1cm} 
d \star \bm{K} = 4 \pi \star \bm{j}, \hspace{1cm}
\bm{j} = C \bm{\xi},
\label{MaxwellEQS}
\end{eqnarray}
where $C$ is a scalar function. Our aim is to solve
Einstein-Maxwell's field equations under these assumptions.
Although the calculations are not very difficult, some
manipulations are required. The details are given in the Appendix .
The solution reads as follows
\begin{eqnarray}
U_1(z) = Q_0 -2 \alpha_i^2 + a_1 \frac{\sinh (2 \beta z)}{2\beta} 
+ \frac{\gamma z}{4} \left [ \left ( 8 \alpha_i - 2 \gamma z \right ) 
\cosh (2 \beta z)  + 3 \gamma  \frac{\sinh (2 \beta z)}{\beta}
\right ] + \nonumber \hspace{7mm}  \\
+ \left ( \nu_0 + 2 \beta^2 \alpha_r^2 + 2 \beta^2 \alpha_i^2 \right)
 \frac{1 - \cosh (2 \beta z) }{2 \beta^2}
- \frac{\mu_0}{2 \beta^2} \left [
\frac{\cosh (2 \beta z) -1}{\beta^2} - \frac{z \sinh (2 \beta z)}{\beta}
\right ] , \\ 
V_1(y) = Q_0 + 2 \alpha_r^2 + a_2 \frac{\sin (2 \beta y)}{2\beta} 
+ \frac{\gamma y}{4} \left [ \left ( 2 \gamma y - 8 \alpha_r \right ) 
\cos (2 \beta y) - 3 \gamma  \frac{\sin (2 \beta y)}{\beta} \right ] +
\nonumber \hspace{13mm} \\
+ \left (\nu_0 - 2 \beta^2 \alpha_r^2 -2 \beta^2  \alpha_i^2 \right )
 \frac{1 - \cos (2 \beta y) }{2 \beta^2}
+ \frac{\mu_0}{2 \beta^2} \left [
\frac{1 - \cos (2 \beta y)}{\beta^2} - \frac{y \sin (2 \beta y)}{\beta}
\right ],
\label{U1V1}
\end{eqnarray}
where $\beta$, $Q_0$, $\mu_0$, $a_1$, $a_2$, $\nu_0$, $\alpha_r$, $\alpha_i$ 
and $\gamma$ are arbitrary constants. These symbols have been
chosen so that the uncharged subcase (i.e. the Wahlquist metric)
can be directly obtained just by setting $\gamma =
\alpha_r = \alpha_i =0$.
Thus, the Wahlquist-Newman family of metrics contains three
more essential parameters than the Wahlquist family. It is worth
pointing out that Kerr-Newman-de Sitter has only one additional
parameter with respect to the Kerr-de Sitter metric (i.e.
the charge of the black hole).  The difference comes from the
fact that, in our case, a non-vanishing charge
current $\vec{j}$ is allowed. The
electromagnetic field $\bm{K}$ of the Wahlquist-Newman spacetime is
\begin{eqnarray*}
\bm{K} = X_r \theta^0 \wedge dy - X_i \theta^1 \wedge dz,
\end{eqnarray*}
where $\theta^0 = d \tau - v_1 d\sigma$, 
$\theta^1 =  d\tau + v_2 d\sigma$,  and  the functions $X_r$ and $X_i$ 
are the real and imaginary parts of the complex function $X = X_r + i X_i$
given by
\begin{eqnarray*}
X = \frac{1}{\left (v_1 + v_2 \right)^2}
\left [ \frac{1- \cos (2 \beta y) \cosh (2 \beta z)}{\beta^2} - i
\frac{\sin (2 \beta y) \sinh (2 \beta z)}{\beta^2} \right ] \times \\
\left [ \alpha_r - \frac{\gamma y}{2} +
 \frac{\gamma}{4} \frac{\sin (2 \beta y)}{\beta} \cosh (2\beta z)
 +  i \left ( \alpha_i - \frac{\gamma z}{2} +
 \frac{\gamma}{4} \frac{\sinh (2 \beta z)}{\beta} \cos (2\beta y)
\right ) \right ]. 
\end{eqnarray*}
The charge current of this electromagnetic field is
\begin{eqnarray}
\vec{j} = \frac{\gamma \beta^2}{2 \pi} \frac{\partial}{\partial \tau}.
\label{j}
\end{eqnarray}
Thus, the constant $\gamma$ is directly related to the charge of the
particles in the fluid. Notice that the value $\gamma=0$
(i.e. uncharged particles) is perfectly possible. In that case, 
the source of the electromagnetic field
lies in the singularity $v_1 + v_2 =0 \Leftrightarrow z=0, y =
n \pi/\beta, n \in \Bbb{Z}$, analogously as in the Kerr-Newman-de Sitter
metric. When $\gamma=0$, the electromagnetic field is described by
the two constants $\alpha_r$ and $\alpha_i$ but only the
combination $\alpha_r^2 + \alpha_i^2$ appears in the metric. This 
reflects the well-known electromagnetic duality symmetry  
of the source-free Einstein-Maxwell field
equations. Thus, for uncharged
particles the Wahlquist-Newman family
adds only one parameter to the Wahlquist family, exactly the same as
in the  Kerr-Newman-de Sitter case.

Regarding the perfect fluid, its velocity
is, by assumption, proportional to  $\partial_{\tau}$ so only
the energy-density $\rho$ and pressure $p$ remain to be given. They can
be directly obtained from the expressions
\begin{eqnarray}
\frac{\rho + 3 p}{2} = \mu_0 + \beta^2 \gamma^2 + \beta \gamma
\frac{\left ( 2 \alpha_r - \gamma y \right)
\sin (2 \beta y) + \left ( 2 \alpha_i - \gamma z \right) \sinh (2 \beta z)}{
v_1 + v_2}, \nonumber \\
\rho + p  = 2 \beta^2 \lambda, \hspace{45mm}
\label{rhop}
\end{eqnarray}
where $\lambda = (V_1 - U_1)(v_1 + v_2) =
- \xi^{\alpha} \xi_{\alpha}$ is minus the squared norm of the Killing
vector. When the particles are uncharged ($\gamma =0$) the perfect fluid
satisfies $\rho + 3 p = 2 \mu_0$
as in the Wahlquist family. When $\gamma \neq 0$, there is no
functional relation between $\rho$ and $p$ and therefore no barotropic
equation of state. Thus, the 
presence of an electric charge in the particles seems to change the
thermodynamic properties of the perfect fluid (this cannot
be made certain until a proper thermodynamic analysis is done).

From (\ref{U1V1}) we observe that both functions $V_1$ and $U_1$
have a smooth limit $\beta \rightarrow 0$ (the integration constants were
chosen carefully so that this property holds). The expressions for the
density and pressure (\ref{rhop}) shows that $\beta =0$ corresponds to
having no perfect fluid but rather a cosmological constant 
with value $\Lambda = -\mu_0$. The electromagnetic field in this
case is source-free, as it should be because no matter
is present. The explicit form
for $v_1$, $v_2$, $U_1$ and $V_1$ in the limit $\beta 
= 0$ is, after redefining $a_1$, $a_2$ and $\nu_0$ so that
the constant $\gamma$ disappears (no trace of $\gamma$ can be left in
this case because the charge current vanishes)
\begin{eqnarray}
v_1 = z^2, \hspace{5mm} v_2 = y^2, \hspace{5mm}
U_1 = Q_0 - 2 \alpha_i^2 + a_1 z
- \nu_0  z^2 + \frac{\mu_0}{3} z^4, \nonumber \\
V_1 = Q_0 + 2 \alpha_r^2 + a_2  y
+ \nu_0 y^2 + \frac{\mu_0}{3} y^4.
\hspace{2cm}
\label{chargedDem}
\end{eqnarray}
The metric (\ref{Wahlgen}) 
with the functions (\ref{chargedDem}) is the Pleba\'nski limit of the
rotating $C$-metric, as expected, and therefore it contains the
Kerr-Newman-de Sitter metric as a particular
case. 

Hence, metric (\ref{Wahlgen}) contains both the Wahlquist and
Kerr-Newman-de Sitter metrics. Furthermore, a simple calculation shows
that the geometric relationship (\ref{weylSi}) is also satisfied
by this metric. Since the self-dual electromagnetic field is 
proportional to the Killing form of $\vec{\xi}$ by
construction, we conclude that (\ref{Wahlgen}) is the
Wahlquist-Newman metric we are seeking (this completes Table 1).
This family of metrics contains eight arbitrary parameters
(or nine if we count $\beta$).

Finally, we can now see why both functions $U_1$ and $V_1$
had to be modified instead of only one as in 
Kerr-Newman-de Sitter. In the cosmological constant case, both
functions get modified by the inclusion
of an electromagnetic field (see (\ref{chargedDem}).
However, a redefinition of $Q_0$ can be used to compensate
one of the
changes. In the perfect-fluid case, the modifications are more complicated and
cannot be reabsorbed by redefinitions of constants.

\section{Extension of the Wahlquist-Newman so\-lu\-tion \\ and static limit}

The metric as written in (\ref{Wahlgen}) does not have an obvious static
limit. Analyzing whether such a limit exists is relevant because
the Wahlquist metric has an interesting, spherically symmetric
static limit, namely the Whittaker
solution \cite{Whit} which represents an isolated fluid ball in equilibrium.
Moreover, the static limit of the Pleba\'nski metric has interesting subcases,
like  the fundamental Schwarzschild-de Sitter-Reissner-Nordstr\"om
metric or the so-called rotating topological black holes (see e.g.
\cite{Vanzo}). Thus, it is reasonable to expect that the static limit
of the  Wahlquist-Newman spacetime may also have interesting properties.
We devote this section to find this limit.

To do that, the coordinate system in (\ref{Wahlgen}) 
must clearly be changed.
We choose a coordinate system which,
in addition, extends the metric (\ref{Wahlgen}) across its
Killing horizon, which is contained
within the set of points  where the Killing vectors $\partial_{\eta}$ and
$\partial_{\sigma}$ span a null two-plane.
Notice, that this can only happen at points where $\lambda \leq 0$. 
From the perfect-fluid interpretation of 
(\ref{Wahlgen}) this  would seem to be impossible.
However, the energy-momentum tensor of 
(\ref{Wahlgen}) is regular at the points where $\lambda =0$, i.e. at
the ergospheres of the Killing vector $\vec{\xi}$. Indeed, the 
electromagnetic field is easily seen to be regular there and even
though the velocity of the perfect fluid becomes singular where
$\lambda =0$, the combination
$(\rho + p) u_{\alpha} u_{\beta} = \lambda^{-1} (\rho + p)
\xi_{\alpha} \xi_{\beta} =
2 \beta^2 \xi_{\alpha} \xi_{\beta}$ is finite.
Obviously the perfect-fluid interpretation
breaks down at the ergospheres of $\vec{\xi}$ 
but still the spacetime is regular.  This indicates that horizons
may also be present in the Wahlquist-Newman spacetime.
In order to find them, we should
evaluate $N=(\partial_{\tau}, \partial_{\tau})
(\partial_{\sigma},\partial_{\sigma}) - (\partial_{\tau},\partial_{\sigma})^2$
where $(\,\, , \,\,)$ means scalar product with the metric (\ref{Wahlgen}).
A simple calculation gives $N =  V_1 U_1$. Thus, $N$ vanishes at the points
where either $V_1$ or $U_1$ vanish. It is not clear a priori
whether we should 
try to extend the metric  across the hypersurface
$V_1=0$ or across the hypersurface $U_1=0$. We known from 
(\ref{coordchange})  that the coordinate $y$ is radial
and $z$ angular, at least in the
limit $\beta =0$ without electromagnetic
field. Therefore, we choose to extend the
spacetime across $V_1(y)=0$. Let us choose the
region $V_1(y)>0$ and define the following coordinate transformation
\begin{eqnarray*}
v = \tau +  \int{\frac{v_2}{V_1} dy} , \hspace{1cm}
\varphi = -\sigma + \int{\frac{1}{V_1} dy}. 
\end{eqnarray*}
It is easy to check that the metric can be cast into the form
\begin{eqnarray}
ds^2 = - \lambda \left (dv + v_1 d \varphi \right )^2 
+ 2 \left ( dy - U_1 d \varphi \right  ) \left (dv + v_1 d \varphi \right )
+ Q \left( \frac{dz^2}{U_1} +
U_1 d\varphi^2  \right ),
\label{extended}
\end{eqnarray}
where $Q \equiv v_1+v_2$. This metric
is  regular at $V_1(y)=0$ and can therefore be extended.
It is straightforward to check that the hypersurface $y=y_0$ with
$V_1(y_0)=0$ is null and that the Killing vector $-v_2(y_0) \partial_{\tau} +
\partial_{\sigma}$ is also null and tangent to this hypersurface.
Thus, $y=y_0$ is a Killing horizon.
Extensions of spacetimes are not
unique in general. The extension we have performed, however, is 
uniquely determined by the geometric condition (\ref{weylSi})
which still holds in the extended spacetime. Thus, 
this is the natural extension of the Wahlquist-Newman metric
from the geometrical
point of view.  It must be emphasized, however, that this extension
may not be
the most relevant from the physical point of view
because the extended region contains, in addition to an electromagnetic
field, a charged tachyonic fluid, which is rather unphysical.

We can now try to determine the static limit of (\ref{extended}).
From Kerr-de Sitter, we know that some limit $z \rightarrow \mbox{const}$
will be involved. So, we should
avoid using $z$ as a coordinate. We accomplish this as follows.
Let us consider a 
connected two-dimensional manifold $S$ endowed with the metric
\begin{eqnarray}
h = \frac{1}{U_1(z)} dz^2 + U_1(z) d\varphi^2, \label{metricah}
\end{eqnarray}
and volume form $\eta_h = dz \wedge  d\varphi$.
Denote by $\star_{h}$ the Hodge dual in 
$(S,h, \eta_{h})$. We obviously have
$\star_{h} d z = U_1 d \varphi$ and
$d \star_{h} d z =  \frac{d U_1}{d z} \eta_{h}$. Furthermore,
the one-form $\bm{\omega} = -v_1 (z) d \varphi$ on $S$
satisfies $d \bm{\omega} = - \frac{d v_1}{dz} \eta_{h}$.
The scalar curvature of the metric (\ref{metricah}) is easily computed to be
$R(h) = - \frac{d^2 U_1}{d z^2}$.
With these definitions, the metric  (\ref{extended}) can be written as
\begin{eqnarray}
ds^2 = - \lambda \left (dv - \bm{\omega} \right)^2 + 2
\left (dy - \star_h d z \right ) \left (dv - \bm{\omega} \right )
+  Q \, h.
\label{geometricform}
\end{eqnarray}
In this metric,  $z$ need not be a coordinate any longer
and can be regarded just as a real function defined
on $S$. The functions $Q$ and $\lambda$ depend on the spacetime point only
through the values of $y$ and $z$ at that point.
$\vec{\xi} = \partial_v$ is static if
\begin{eqnarray*}
\bm{\xi} \wedge d \bm{\xi} = - V_1 \lambda_{,z} dv \wedge \eta_{h}
+ dy \wedge \left [ \frac{}{} Q \lambda_{\,z} \eta_{h} +
\left (\lambda_{,z} d z + \lambda_{,y} \star_{h} d z \right ) \wedge
\left (\bm{\omega} - dv \right ) \right ]=0,
\end{eqnarray*}
which holds if and only if
$\lambda_{,z} =0$ and $\star_{h} d z =0$. Thus, $z = z_0 = \mbox{const}$
and $\lambda_{,z} |_{z=z_0} =0$. From $\lambda =
(V_1 - U_1)/Q$ 
and (\ref{v1v2}), (\ref{U1V1}) this can only happen
iff $z_0 = 0$ and $a_1 = - 2 \gamma \alpha_i$. 
In that case, the one-form $\bm{\omega}$ and
the scalar curvature of $h$ are
\begin{eqnarray*}
d \bm{\omega} = - \left . \left ( \frac{d v_1}{dz} \right |_{z=0}
\right ) \eta_h = 0,
\hspace{1cm} 
R(h) = - \left .\frac{d^2 U_1}{dz^2} \right |_{z=0} = 2 \left [ 
 \nu_0 + 2 \beta^2 \left (\alpha_r^2 + \alpha_i^2 \right ) - \gamma^2 \right ].
\end{eqnarray*}
Thus, $\bm{\omega}$ is locally exact and can be reabsorbed into
the coordinate $v$. Since $h$ is of constant curvature,
there exist coordinates $x_1$ and
$x_2$ such that
\begin{eqnarray*}
h = B^2 \left [ d{x_1}^2 + \Sigma (x_1,\epsilon ) d{x_2}^2 \right ],
\end{eqnarray*}
where $\Sigma(-1,x_1) = \sinh(x_1)$, $\Sigma(0,x_1) = x_1$ and $
\Sigma(1,x_1) = \sin(x_1)$ and $B \in \Bbb{R}$ satisfies
\begin{eqnarray}
\label{defB}
\epsilon B^{-2} = \nu_0 + 2 \beta^2 \left (\alpha_r^2 + 
\alpha_i^2 \right ) - \gamma^2.
\end{eqnarray}
Inserting this into (\ref{geometricform}) we find that the static limit of the
Wahlquist-Newman metric is
\begin{eqnarray}
ds^2 = - \tilde{\lambda} dv^2 + 2 dy dv + \frac{1 - 
\cos ( 2 \beta y)}{2 \beta^2} B^2 
\left ( d{x_1}^2 + \Sigma (x_1,\epsilon ) d{x_2}^2 \right ),\label{staticmet}
\end{eqnarray}
where $\tilde{\lambda} \equiv \lambda (y,0)$ reads explicitly
\begin{eqnarray*}
\tilde{\lambda} =
\left (\nu_0 - 2 \beta^2 \alpha_r^2 -2 \beta^2  \alpha_i^2 \right )
+ \frac{2 \beta^2}{1 - \cos \left (2 \beta y \right )} \left \{
2 \left (\alpha_r^2 + \alpha_i^2 \right )
+ a_2 \frac{\sin (2 \beta y)}{2\beta} 
+ \nonumber \right .\\
\left . \frac{}{} \frac{\mu_0}{2 \beta^2} \left [
\frac{1 - \cos (2 \beta y)}{\beta^2} - \frac{y \sin (2 \beta y)}{\beta}
\right ] + \frac{\gamma y}{4} \left [ \left ( 2 \gamma y - 8 \alpha_r \right ) 
\cos (2 \beta y) - 3 \gamma  \frac{\sin (2 \beta y)}{\beta} \right ]
\right \}.
\end{eqnarray*}
We call this metric  Whittaker-Reissner-Nordstr\"om metric.
Its energy-momentum tensor is (in the region $\lambda >0$) 
the sum of a perfect-fluid and an electromagnetic field. The density
and pressure of the perfect fluid can be read off from (\ref{rhop})
after inserting $z=0$. The electromagnetic field can be obtained
by performing the coordinate changes we made in get the static limit.
The result is
\begin{eqnarray*}
K = \frac{2 \beta^2}{1 - \cos (2 \beta y)}
\left [2 \alpha_r + \gamma \left (\frac{\sin (2\beta y)}{2 \beta}
- y \right ) \right ] dv \wedge dy - 2 \alpha_i \eta_h,
\end{eqnarray*}
where the two-form $\eta_h$ is $\eta_h = B^2 \Sigma(x_1,\epsilon) \, \,dx_1
\wedge dx_2$. Its charge current is still given by (\ref{j}).
The metric (\ref{staticmet}) is static and spherically
symmetric  as long as $[\nu_0 + 2 \beta^2 (\alpha_r^2
+ \alpha_i^2) - \gamma^2] >0$. When this expression is zero or negative, the
spacetime is plane symmetric and ``hyperbolic'' symmetric respectively.
When the electromagnetic field vanishes and 
$\nu_0 >0$ the metric is the spherically symmetric perfect-fluid found by
Whittaker \cite{Whit}. The limit $\beta \rightarrow 0$ gives the
de Sitter-Reissner-Nordstr\"om metric (when $\epsilon =1$) or its
hyperbolic or plane counterparts. Another
physically relevant subcase of Whittaker-Reissner-Nordstr\"om
is $\alpha_r = \alpha_i = a_2 =0$ and $\nu_0 > \gamma^2$. This
represents a charged fluid ball in equilibrium, with no singularities
inside. 

A thorough investigation of the geometry of the Wahlquist-Newman and
Whittaker-Reissner-Nordstr\"om spacetimes would be of interest. Also,
studying the physical applications of this geometrically privileged metrics
should be a matter of further investigation.

\section*{Acknowledgements}
I would like to thank J.M.M Senovilla for very useful suggestions and
criticisms on a previous version of the paper. I would also
like to thank the Albert Einstein Institute for kind hospitality.

\section*{Appendix}

In this appendix we solve the Einstein-Maxwell equations under the assumptions
described in Sect. \ref{WN}. Let us start be introducing an orthogonal tetrad
\begin{eqnarray}
\theta^0 = d\tau - v_1 d\sigma, \hspace{1cm}
\theta^1 = d\tau + v_2 d\sigma, \hspace{1cm}
\theta^2 = dy, \hspace{1cm}
\theta^3 = dz,
\label{tetrad}
\end{eqnarray}
so that the metric  (\ref{Wahlgen}) takes the form
\begin{eqnarray*}
ds^2 = -\frac{V_1}{v_1 + v_2} {(\theta^0)}^2 +
\frac{U_1}{v_1 + v_2} {(\theta^1)}^2 +
 \frac{v_1 + v_2}{V_1} {(\theta^2)}^2 +
\frac{v_1 + v_2}{U_1} {(\theta^3)}^2. 
\end{eqnarray*}
We take the volume form $\eta=
\theta^0 \wedge \theta^1 \wedge \theta^2 \wedge \theta^3$.
Lowering the indices to  $\vec{\xi} = \partial_{\tau}$ we find
\begin{eqnarray*}
\bm{\xi} = - \frac{V_1}{v_1+v_2} \theta^0 + \frac{U_1}{v_1+v_2} \theta^1.
\end{eqnarray*}
In order to impose $\K_{\alpha\beta} \propto \F_{\alpha\beta}$, we
need to evaluate the Killing form $\cm{F}$ associated to $\vec{\xi}$.
After a simple computation we obtain
\begin{eqnarray}
\cm{F} = \frac{1}{2} \left [ \frac{ {V_1}_{,y} + i {U_1}_{,z}}{v_1 + v_2}
+ \frac{V_1-U_1}{\left (v_1 + v_2 \right )^2} \left ( i {v_1}_{,z}  
- {v_2}_{,y} \right ) \right ] \left ( 
\theta^0 \wedge \theta^2 + i
\theta^1 \wedge \theta^3 \right ), 
\label{KillForm}
\end{eqnarray}
where $Q = v_1 + v_2$.
Thus, two of the three linearly independent (complex) coefficients
of the Killing form $\cm{F}$ are identically zero.
Since the fluid velocity $\vec{u} \propto \vec{\xi}$, the
perfect-fluid part of the energy-momentum tensor reads
\begin{eqnarray*}
T^{pf} = \left ( D \frac{V_1^2}{Q^2} - p \frac{V_1}{Q} \right ) 
{(\theta^0)}^2   - 2 D \frac{V_1 U_1}{Q^2} 
\theta^0 \theta^1 + \left ( D \frac{U_1^2 }{Q^2} + p \frac{U_1 }{Q} \right )
{(\theta^1)}^2 +
p Q \left ( \frac{{(\theta^2)}^2}{V_1} +
\frac{{(\theta^3)}^2}{U_1} \right ),
\end{eqnarray*}
where $p$ is the pressure and the density $\rho$ is obtained from the scalar
$D$ by $\rho + p = Q^{-1} D \left (U_1 - V_1 \right )$.
The electromagnetic field $\bm{K}$ is required to satisfy
\begin{eqnarray*}
\cm{K} = { X} \left [ \theta^0 \wedge \theta^2 + i \,\, \theta^1 \wedge
\theta^3 \right ],
\end{eqnarray*}
where ${X}$ is a complex scalar function. Hence, the 
electromagnetic energy-momentum tensor 
$T^{em}_{\mu\nu} = (1/4) \cm{K}_{\mu\alpha}
\cm{\ov{K}}^{\,\,\,\,\alpha}_{\nu}$ takes the 
diagonal form
\begin{eqnarray*}
T^{em} = \frac{1}{2} {X} \ov{{X}} \left [
\frac{V_1}{Q} {(\theta^0)}^2+
\frac{U_1}{Q} {(\theta^1)}^2-
\frac{Q}{V_1} {(\theta^2)}^2+
\frac{Q}{U_1} {(\theta^3)}^2 \right ].
\end{eqnarray*}
Using units in which $8 \pi G = c =1$ and denoting by $G_{\alpha\beta}$
the Einstein tensor of (\ref{Wahlgen}), the Einstein equations $G_{\mu\nu}=
T^{em}_{\mu\nu}+ T^{pf}_{\mu\nu}$ become
\begin{eqnarray}
G_{00} + \frac{V_1}{U_1} G_{01} + \frac{V_1^2}{Q^2} G_{22} = 0, 
\hspace{1cm}
G_{11} + \frac{U_1}{V_1} G_{01} - \frac{U_1^2}{Q^2} G_{33} = 0, 
\label{Ein1} \\
D = - \frac{ G_{01} Q^2}{V_1 U_1}, \hspace{1cm}
p = \frac{V_1 G_{22}}{2Q} + \frac{U_1 G_{33}}{2 Q}, 
\hspace{18mm} \label{Ein2} \\
X \ov {X} =  \frac{U_1 G_{33}}{Q} - \frac{V_1 G_{22}}{Q}.
\label{Ein3} \hspace{35mm}
\end{eqnarray}
The two equations (\ref{Ein1}) are identically satisfied 
by the metric (\ref{Wahlgen}). Actually, it can be proven that allowing
$v_1(y)$ and $v_2(z)$ to be arbitrary, the two equations
(\ref{Ein1}) force them to be (\ref{v1v2}).
Thus, our assumption that $v_1$ and $v_2$ remain unchanged 
implies no loss of generality. The two equations in (\ref{Ein2}) 
can be regarded as defining expressions for $\rho$ and $p$
(we do not impose any equation of state for the
perfect fluid a priori). Equation (\ref{Ein3}) needs to be solved in
combination with the Maxwell's equations (\ref{MaxwellEQS}),
which we now analyze.
The electromagnetic field is required to be
Lie constant along the Killing vector fields $\partial_{\tau}$ and
$\partial_{\sigma}$. Thus ${X} = { X}(y,z)$. 
In our case, it is simpler to solve Maxwell's equations by looking for an 
electromagnetic potential $\bm{A}$ satisfying $d \bm{A} = \bm{K}$. 
Since $\bm{A}$ can be chosen to be Lie constant along the Killing vectors,
we can write
\begin{eqnarray*}
\bm{A} = A_{0}(y,z) \theta^0 + 
A_{1}(y,z) \theta^1+ 
A_{2}(y,z) \theta^2 + 
A_{3}(y,z) \theta^3,
\end{eqnarray*}
so that its exterior derivative takes the form
\begin{eqnarray}
d \bm{A} = \left [ - \partial_y A_0 + \frac{\sin (2 \beta y)}{\beta} 
\frac{A_1}{Q} \right ] \theta^0 \wedge \theta^2 - \left [
\partial_z A_0 + \frac{\sinh (2 \beta z)}{\beta} \frac{A_0}{Q} \right ]
\theta^0 \wedge \theta^3 - \left [ \frac{}{} \partial_y A_1 + 
\right . \nonumber \\
\left . 
\frac{\sin (2 \beta y)}{\beta} \frac{A_1}{Q} \right ] \theta^1 \wedge \theta^2
+ \left [ - \partial_z A_1 + \frac{\sinh (2 \beta z)}{\beta} \frac{A_0}{Q}
\right ] \theta^1 \wedge \theta^3 + \left [ \partial_y A_3 - \partial_z
A_2 \right ] \theta^2 \wedge \theta^3.
\label{exprA}
\end{eqnarray}
Decomposing ${ X}$ into its real and imaginary parts ${ X}
 = X_r + i X_i$, the electromagnetic field reads
 $\bm{K} = X_r \theta^0 \wedge \theta^2- X_i \theta^1 \wedge \theta^3$.
Imposing now $d \bm{A} = \bm{K}$ we obtain, first of all, that
the coefficient in $\theta^2 \wedge \theta^3$ must vanish. Thus 
$A_2 \theta^2 + A_3 \theta^3$ is closed and can be redefined away by a
a gauge transformation. So, we can put $A_2 = A_3 =0$.
The vanishing of the coefficients 
in $\theta^0 \wedge \theta^3$ and
$\theta^1 \wedge \theta^2$  in (\ref{exprA}) implies
$A_0 = \tilde{A}_0(y)/Q$ and $A_1 = \tilde{A}_1(z)/Q$.
The remaining components of $d \bm{A} = \bm{K}$ give expressions
for $X_r$ and $X_i$ in terms of $\tilde{A}_0$ and $\tilde{A}_1$
and their derivatives. A convenient way of writing them is
\begin{eqnarray}
X_r = -  \partial_y  \left [
\frac{\tilde{A}_0 + \tilde{A}_{1} }{v_1 + v_2 } \right ], 
\hspace{1cm}
X_i = \partial_z  \left [
\frac{\tilde{A}_0 + \tilde{A}_{1} }{v_1 + v_2 } \right ].
\label{XiXr}
\end{eqnarray}
We turn now into the equation $d \star \bm{K} = 4 \pi \star \bm{j}$, which 
after using the form of $\bm{K}$ and $\bm{j}$ reads
\begin{eqnarray}
\left ( X_{i,z} + X_i \frac{v_{1,z}}{Q} + X_r \frac{v_{2,y}}{Q} \right )
\theta^0 \wedge \theta^2 \wedge \theta^3 +
\left ( - X_{r,y} - X_i \frac{v_{1,z}}{Q} - X_r \frac{v_{2,y}}{Q} \right )
\theta^1 \wedge \theta^2 \wedge \theta^3 \nonumber \\
= 4 \pi C \left ( \theta^1 \wedge \theta^2 \wedge \theta^3 -
\theta^0 \wedge \theta^2 \wedge \theta^3 \right ).
\label{equationC}
\end{eqnarray}
This implies $X_{r,y} - X_{i,z}=0$, or using (\ref{XiXr}),
\begin{eqnarray}
\left ( \partial_{yy} + \partial_{zz} \right ) \left [
\frac{\tilde{A}_0 + \tilde{A}_1}{Q} \right ]=0.
\label{secondorder}
\end{eqnarray}
Defining the complex variable $\zeta = y + i z$, the general
solution of (\ref{secondorder}) is 
$\tilde{A}_0 + \tilde{A}_1 = Q \cdot [ g(\zeta) + \ov{g(\zeta)} ]$, where
$g$ is a holomorphic function of $\zeta$. In terms of 
$\zeta$, the function $Q = v_1 + v_2$ becomes simply $Q = \beta^{-2}
\sin(\beta \zeta) \sin (\beta \ov{\zeta})$. It remains to impose 
that $\tilde{A}_0$ and $\tilde{A}_1$ depend only on $y$ and $z$ respectively,
or equivalently $\left (\partial_{\zeta \zeta} -  \partial_{\,\ov{\zeta}
\ov{\zeta}} \right ) \left (
\tilde{A}_0 + \tilde{A}_1 \right )=0$. This implies the following
equation for $g$,
\begin{eqnarray*}
g_{,\zeta \zeta} + 2 \beta \frac{\cos(\beta \zeta)}{\sin(\beta \zeta) }
g_{\zeta}
= \ov{g}_{,\ov{\zeta} \ov{\zeta}} + 2 
\beta \frac{\cos(\beta \ov{\zeta})}{\sin(\beta \ov{\zeta}) }
\ov{g}_{\ov{\zeta}} \,\,.
\end{eqnarray*}
Thus, there exits a real constant $\beta^2 \gamma$ such that
each term of this equation equals $\beta^2 \gamma$. The resulting
ODE can be integrated once to give
(after choosing the integration
constant so that the limit $\beta 
\rightarrow 0$ exists)
\begin{eqnarray}
g_{\zeta} = \frac{\beta^2}{\sin^2(\beta \zeta)} \left [
- \alpha + \gamma \left (\frac{\zeta}{2}
- \frac{\sin(\beta \zeta) \cos (\beta \zeta) }{2 \beta} \right )
\right ], \label{gxi}
\end{eqnarray}
where $\alpha$ is an arbitrary complex constant.
From this expression we could easily integrate $g (\zeta)$ and
obtain $\bm{A}$. However, to obtain $\bm{K}$ we only need to determine
$X$,
\begin{eqnarray*}
X = X_r + i X_i = \left ( - \frac{\partial}{\partial y} +
i \frac{\partial}{\partial z} \right ) \left (
\frac{\tilde{A}_0 + \tilde{A}_1}{Q} \right )
= -2 \frac{\partial}{\partial \zeta} \left ( g(\zeta) +
\ov{g} (\ov{\zeta} ) \right ) =  -2 g_{,\zeta}.
\end{eqnarray*}
The scalar $C$ can now be read off from (\ref{equationC}), the
result being $C = \gamma \beta^2/2 \pi$. We can now solve the 
Einstein field equation (\ref{Ein3}). First, we need to evaluate
$X \ov{X}$. Decomposing $\alpha$ into its real and imaginary
parts as $\alpha = \alpha_r + i \alpha_i$, and using (\ref{gxi})
we get $X \ov{X} = 4 Q^{-1} Y \ov{Y}$, where
\begin{eqnarray*}
Y = - \alpha_r + \frac{\gamma y}{2} -
 \frac{\gamma}{4} \frac{\sin (2 \beta y)}{\beta} \cosh (2\beta z)
 + i \left ( - \alpha_i + \frac{\gamma z}{2} -
 \frac{\gamma}{4} \frac{\sinh (2 \beta z)}{\beta} \cos (2\beta y)
\right ).
\end{eqnarray*}
Einstein's equation (\ref{Ein3}) reads, after dropping a factor $Q =
v_1 + v_2$,
\begin{eqnarray}
\left ( v_1 + v_2 \right ) \left ( U_1 G_{33} - V_1 G_{22} \right
) - 4 Y \ov{Y} =0,
\label{Ein4}
\end{eqnarray}
which is a rather long equation involving
the functions $V_1$, $U_1$ and their derivatives.
Since they are functions of different variables, a reasonable strategy is to
try and separate this equation. This can be accomplished after taking the
partial derivative of (\ref{Ein4}) with respect to $y$ and $z$. The resulting
expression separates nicely into the form
\begin{eqnarray*}
\frac{\beta}{\sin (2 \beta y) } \left [
V_{1,yyy} + 4 \beta^2 V_{1,y} + 8 \beta^2 \gamma \cos (2 \beta y)
\left ( \gamma y - 2 \alpha_r \right ) \right ]
= \hspace{4cm}\\
\hspace{2cm}
\frac{\beta}{\sinh (2 \beta z) } \left [
U_{1,zzz} - 4 \beta^2 U_{1,z} + 8 \beta^2 \gamma \cosh (2 \beta z)
\left ( \gamma z - 2 \alpha_i \right ) \right ] = 4 \mu_0,
\end{eqnarray*}
where $\mu_0$ is the separation constant. Thus, we are faced with two
linear, third order ordinary differential equations. Their solution
can be explicitly written down in the following form, after choosing carefully
the integration constants so that
the limit $\beta \rightarrow 0$ exists, 
\begin{eqnarray*}
U_1 = L_1 + a_1 \frac{\sinh (2 \beta z)}{2\beta} 
+ \frac{\gamma z}{4} \left [ \left ( 8 \alpha_i - 2 \gamma z \right )
 \cosh (2 \beta z)  + 3 \gamma  \frac{\sinh (2 \beta z)}{\beta}
\right ] + \hspace{1cm} \\
+ S_1 \frac{\cosh (2 \beta z) - 1 }{2 \beta^2} 
- \frac{\mu_0}{2 \beta^2} \left [
\frac{\cosh (2 \beta z) -1}{\beta^2} - \frac{z \sinh (2 \beta z)}{\beta}
\right ], \\
V_1 = L_0 + a_2 \frac{\sin (2 \beta y)}{2\beta} +
\frac{\gamma y}{4} \left [ \left ( 2 \gamma y - 8 \alpha_r \right )
\cos (2 \beta y)
- 3 \gamma  \frac{\sin (2 \beta y)}{\beta}
\right ] + \hspace{1cm} \\
 + S_0 \frac{1 - \cos (2 \beta y) }{2 \beta^2} 
\frac{\mu_0}{2 \beta^2} \left [
\frac{1 - \cos (2 \beta y)}{\beta^2} - \frac{y \sin (2 \beta y)}{\beta}
\right ], \\
\end{eqnarray*}
where $L_0$, $L_1$, $S_0$, $S_1$, $a_1$ and $a_2$ are integration
constants. Inserting these expressions back into the Einstein equation
(\ref{Ein4}), we find that the equation is satisfied if and only if 
$S_1 + S_0 = - 4 \beta^2 \left (\alpha_r^2 + \alpha_i^2
 \right )$ and $L_1 = L_0 - 2  \left (\alpha_r^2 + \alpha_i^2 \right)$.
By redefining $S_0$, $S_1$ and $L_0$ in a trivial way we obtain
the form for $U_1$ and $V_1$ given in (\ref{U1V1}).

\end{document}